\documentclass[aps,groupedaddress,showpacs,floatfix]{revtex4}
\usepackage{amssymb}
\usepackage{amsmath}
\usepackage{graphicx}
\DeclareMathOperator{\Ai}{Ai}
\begin{document}

\title{Bethe anzats derivation of the Tracy-Widom distribution for
one-dimensional directed polymers}

\author{Victor Dotsenko }

\affiliation{LPTMC, Universit\'e Paris VI, 75252 Paris, France}

\affiliation{L.D.\ Landau Institute for Theoretical Physics,
   119334 Moscow, Russia}

\date{\today}

\begin{abstract}

The distribution function of the free energy fluctuations 
in one-dimensional directed polymers with 
$\delta$-correlated random potential is studied by mapping the
replicated problem to a many body quantum boson system with attractive
interactions.  Performing the summation over the entire spectrum of excited
states the problem is reduced to the Fredholm determinant with
the Airy kernel which is known to yield the  Tracy-Widom distribution.

\end{abstract}


\maketitle

\medskip

\section{Introduction}

Directed polymers in a quenched random potential have 
been the subject of intense investigations during the past two
decades (see e.g. \cite{hh_zhang_95}). In the most simple one-dimensional case
we deal with an elastic string  directed along the $\tau$-axis 
within an interval $[0,L]$. Randomness enters the problem 
through a disorder potential $V[\phi(\tau),\tau]$, which competes against 
the elastic energy.  The problem is defined by the Hamiltonian
\begin{equation}
   \label{bas1-1}
   H[\phi(\tau), V] = \int_{0}^{L} d\tau
   \Bigl\{\frac{1}{2} \bigl[\partial_\tau \phi(\tau)\bigr]^2 
   + V[\phi(\tau),\tau]\Bigr\};
\end{equation}
where in the simplest case the disorder potential $V[\phi,\tau]$ 
is Gaussian distributed with a zero mean $\overline{V(\phi,\tau)}=0$ 
and the $\delta$-correlations: 
\begin{equation}
   \label{bas1-2}
{\overline{V(\phi,\tau)V(\phi',\tau')}} = u \delta(\tau-\tau') \delta(\phi-\phi')
\end{equation}
Here the parameter $u$ describes the strength of the disorder.
Historically, the problem of central interest was the scaling behavior of the 
polymer mean squared displacement  which in the thermodynamic limit
($L \to \infty$) is believed to have a universal scaling form
$\overline{\langle\phi^{2}\rangle}(L) \propto L^{2\zeta} $
(where $\langle \dots \rangle$ and $\overline{(\dots)}$ denote 
the thermal and the disorder averages), with $\zeta=2/3$, the so-called wandering exponent. 
More general problem for all directed polymer systems of the type, Eq.(\ref{bas1-1}),
is the statistical properties  of their free energy fluctuations. 
Besides the usual extensive (linear in $L$) self-averaging part $f_{0} L$ 
(where $f_{0}$ is the linear free energy density),
the total free energy $F$ of such systems contains disorder dependent 
fluctuating contribution $\tilde{F}$, which is characterized by
non-trivial scaling in $L$. It is generally believed  that in the limit of large $L$ 
the typical value of the free energy fluctuations scales with $L$ as
$\tilde{F}  \propto L^{1/3}$ (see e.g. \cite{hhf_85,numer1,numer2,kardar_87}
In other words, in the limit of large $L$ the total (random) free energy of the system 
can be represented as
\begin{equation}
\label{bas1-5}
F \; = \; f_{0} L \; + \; c \, L^{1/3}\; f
\end{equation}
where $c$ is the parameter, which depends on the temperature and the strength of disorder, and 
$f$ is the random quantity which in the thermodynamic limit $L\to\infty$ 
is described by a non-trivial universal
distribution function $P_{*}(f)$. The derivation of this function for the system with $\delta$-correlated
random potential, Eqs.(\ref{bas1-1})-(\ref{bas1-2}) is the central issue of the present work.

 For the string with the zero boundary conditions
at $\tau=0$ and at $\tau=L$ the partition function of a given sample is
\begin{equation}
\label{bas1-6}
   Z[V] = \int_{\phi(0)=0}^{\phi(L)=0} 
              {\cal D} [\phi(\tau)]  \;  \mbox{\Large e}^{-\beta H[\phi,V]}
\end{equation}
where $\beta$ denotes the inverse temperature. On the other hand,
the partition function is related to the total free energy $F[V]$ via
$Z[V] = \exp( -\beta  F[V])$.
The free energy $F[V]$  is defined for a specific 
realization of the random potential $V$ and thus represent a random variable. 
Taking the $N$-th power of both sides of this relation 
and performing the averaging over the random potential $V$ we obtain
\begin{equation}
\label{bas1-8}
\overline{Z^{N}[V]} \equiv Z[N,L] = \overline{\exp( -\beta N F[V]) }
\end{equation}
where the quantity in the lhs of the above equation is called the replica partition function.
Substituting here $F = f_{0} L +  c \, L^{1/3}\; f$, 
 and redefining $Z[N,L] = \tilde{Z}[N,L] \;\exp\{-\beta N f_{0} L\} $
we get
\begin{equation}
\label{bas1-11}
\tilde{Z}[N,L] = \overline{\exp( -\lambda N f) }
\end{equation}
where $\lambda = \beta c L^{1/3}$.
The averaging in the rhs of the above equation can be represented in terms of the 
distribution function $P_{L}(f)$ (which depends on the system size $L$). 
In this way we arrive to the following general relation 
between the replica partition function $\tilde{Z}[N,L]$ and the distribution function 
of the free energy fluctuations $P_{L}(f)$:
\begin{equation}
\label{bas1-12}
   \tilde{Z}[N,L] \; =\;
           \int_{-\infty}^{+\infty} df \, P_{L} (f) \;  
           \mbox{\Large e}^{ -\lambda N   \, f}
\end{equation}
Of course, the most interesting object is the thermodynamic limit 
distribution function $P_{*}(f) = \lim_{L\to\infty} P_{L} (f)$ which is expected to be 
the universal quantity. The above equation is the bilateral Laplace transform of  the function $P_{L}(f)$,
and at least formally it allows to restore this function via inverse Laplace transform 
 of the replica partition function $\tilde{Z}[N,L]$. In order to do so one has to compute 
$\tilde{Z}[N,L]$ for an  arbitrary 
integer $N$ and then perform  analytical continuation of this function 
from integer to arbitrary complex values of $N$. 
In Kardar's original solution \cite{kardar_87}, after
mapping the replicated problem to interacting quantum bosons, one arrives at
the replica partition function for positive integer parameters $N > 1$.
Assuming a large $L \to \infty$ limit, one is tempted to approximate the
result by the ground state contribution only, as for any $N > 1$ the
contributions of excited states are exponentially small for $L\to\infty$.
However, in the analytic continuation for arbitrary complex $N$ the
contributions which are exponentially small at positive integer $N > 1$ can
become essential in the region $N\to 0$, which  defines the function $P(f)$
(in other word, the problem is that the two limits $L\to\infty$ and $N\to 0$ 
do not commute \cite{Medina_93,dirpoly}).
Thus, it is the neglection of the excited states which is the origin of non-physical nature 
of the obtained solution. 

In the my recent paper \cite{BAS} an attempt has been made to 
derive the free energy distribution function via the calculation
of the replica partition function $Z[N,L]$ in terms of the 
Bethe-Ansatz solution for quantum bosons with attractive $\delta$-interactions
which involved the summation over the {\it entire spectrum} of
exited states. Unfortunately, the attempt has failed because 
on one hand, the calculations contained a kind of a hidden "uncontrolled  approximation",
and on the other hand, 
the analytic continuation of obtained $Z(N,L)$ was found  to be ambiguous.

It turns out that it is possible to bypass the problem of the analytic continuation if 
 instead of the
distribution function itself one would study its integral
representation
\begin{equation}
 \label{tw1}
W(x) \; = \; \int_{x}^{\infty} \; df \; P_{*}(f)  
\end{equation}
which gives the probability to find the fluctuation $f$ bigger that a given value $x$.
Formally the function $W(x)$ can be defined as follows:
\begin{equation}
 \label{tw2}
W(x) \; = \; \lim_{\lambda\to\infty} \sum_{N=0}^{\infty} \frac{(-1)^{N}}{N!} 
\exp(\lambda N x) \; \overline{\tilde{Z}^{N}} \; = \; 
\lim_{\lambda\to\infty} \overline{\exp\bigl[-\exp\bigl(\lambda (x-f)\bigr)\bigr]} \; = \; 
\overline{\theta(f-x)}
\end{equation}
On the other hand, in terms of the replica approach the function $W(x)$
is given by the series
\begin{equation}
 \label{tw3}
W(x) \; = \; \lim_{\lambda\to\infty} \sum_{N=0}^{\infty} \frac{(-1)^{N}}{N!} 
\exp(\lambda N x) \; \tilde{Z}[N,L]
\end{equation}
In the present paper the replica partition function $\tilde{Z}[N,L]$
will be calculated (again) by mapping the
replicated problem to the $N$-particle quantum boson system with attractive
interactions.  Performing the summation over the entire spectrum of excited
states the summation of the series, eq.(\ref{tw3}), is reduced to the Fredholm 
determinant with the so called Airy kernel which is known to yield the  
Tracy-Widom distribution. Originally this distribution has been derived in the context 
of the statistical properties of the Gaussian Unitary Ensemble \cite{Tracy-Widom}
while at present it is well established to 
describe the statistics of fluctuations in various random systems
\cite{PNG_Spohn,LIS,LCS,oriented_boiling,ballistic_decomposition,DP_johansson}
which are widely believed to belong to the same universality class as the
present model \cite{Derrida1,KK,Prahofer-Spohn}. 
While this manuscript was in course of preparation 
 I have learned that the exactly {\it the same} result for the system 
considered in this paper has been independently derived by P.Calabrese, P.Le Doussal and A.Rosso
\cite{LeDoussal}.

%%%%%%%%%%%%%%%%%%%%%%%%%%%%%%%%%%%%%%%%%%%%%%%%%%%%%%%%%%%%%%%%%%%%%%%%%%%%%%%%%
%%%%%%%%%%%%%%%%%%%%%%%%%%%%%%%%%%%%%%%%%%%%%%%%%%%%%%%%%%%%%%%%%%%%%%%%%%%%%%%%%

\vspace{5mm}

Performing simple Gaussian average over the random potential, eq.(\ref{bas1-2}),
for the replica partition function, Eq.(\ref{bas1-8}), 
we obtain the standard expression
\begin{equation}
   \label{bas2-5}
   Z(N,L) = \prod_{a=1}^{N} \int_{\phi_{a}(0)=0}^{\phi_{a}(L)=0} 
   {\cal D} \phi_{a}(\tau) \;
   \mbox{\Large e}^{-\beta H_{N}[{\boldsymbol \phi}] }
\end{equation}
where
\begin{equation}
\label{bas2-4}
   H_{N}[{\boldsymbol \phi}] =  
   \frac{1}{2} \int_{0}^{L} d\tau \Biggl(
   \sum_{a=1}^{N} \bigl[\partial_\tau\phi_{a}(\tau)\bigr]^2 
   - \beta u \sum_{a\not= b}^{N} 
   \delta\bigl[\phi_{a}(\tau)-\phi_{b}(\tau)\bigr] \Biggr)
\end{equation}
is the $N$-component scalar field replica Hamiltonian and 
${\boldsymbol \phi} \equiv \{\phi_{1},\dots, \phi_{N}\}$.

According to the above definition this partition function describe the statistics
of $N$ $\delta$-interacting (attracting) trajectories $\phi_{a}(\tau)$ all starting 
(at $\tau=0$) and ending (at $\tau=L$) at zero.
In order to map the problem to one-dimensional quantum bosons, 
 let us introduce more general object
\begin{equation}
   \label{bas2-6}
   \Psi({\bf x}; t) = 
\prod_{a=1}^{N} \int_{\phi_a(0)=0}^{\phi_a(t)=x_a} {\cal D} \phi_a(\tau)
  \;  \mbox{\Large e}^{-\beta H_{N} [{\boldsymbol \phi}]}
\end{equation}
which describes $N$ trajectories $\phi_{a}(\tau)$ all starting at zero ($\phi_{a}(0) = 0$),
but ending at $\tau = t$ in arbitrary given points $\{x_{1}, ..., x_{N}\}$.
One can easily show that instead of using the path integral, $\Psi({\bf x}; t)$
may be obtained as the solution of the  linear differential equation
\begin{equation}
   \label{bas2-7}
\partial_t \Psi({\bf x}; t) \; = \;
\frac{1}{2\beta}\sum_{a=1}^{N}\partial_{x_a}^2 \Psi({\bf x}; t)
  \; + \; \frac{1}{2}\beta^2 u \sum_{a\not=b}^{N} \delta(x_a-x_b) \Psi({\bf x}; t)
\end{equation}
with the initial condition $\Psi({\bf x}; 0) = \Pi_{a=1}^{N} \delta(x_a)$.
This  is nothing else but the imaginary-time
Schr\"odinger equation which describes  $N$ bose-particles of mass $\beta $ interacting via
the {\it attractive} two-body potential $-\beta^2 u \delta(x)$. 
The original replica partition function, Eq.(\ref{bas2-5}), then is obtained via a particular
choice of the final-point coordinates, $Z(N,L) = \Psi({\bf 0};L)$

The spectrum and some properties of the eigenfunctions
for attractive one-dimensional quantum bosons have been derived by McGuire
\cite{McGuire} and by Yang \cite{Yang} (see also Ref.\ \cite{Takahashi,Calabrese}).
A generic eigenstate of this system consists of $M$  ($1 \leq M \leq N$)  
"clusters" $\{\Omega_{\alpha}\}$ $(\alpha = 1,...,M)$ of bound particles. 
Each cluster is characterized by the momentum $q_{\alpha}$
of its center of mass motion, and by the number $n_{\alpha}$ of particles contained in it
(such that $\sum_{\alpha=1}^{M} n_{\alpha} = N$).  
Correspondingly, the eigenfunction $\Psi_{\bf q, n}^{(M)}(x_1,...,x_N)$ of such state
is characterized by $M$ continuous parameters ${\bf q} = (q_{1}, ..., q_{M})$  and $M$ 
integer parameters ${\bf n} = (n_{1}, ..., n_{M})$ (detailed structure and the 
properties of these wave functions are described in Ref.\cite{BAS}). The energy spectrum of this 
state is
\begin{equation}
\label{bas3-18}
E_{M}({\bf q,n}) \; = \;
\frac{1}{2\beta} \sum_{\alpha=1}^{M} \; \sum_{r=1}^{n_{\alpha}} (q^{\alpha}_{k})^{2} 
\; = \; \frac{1}{2\beta} \sum_{\alpha=1}^{M} \; n_{\alpha} q_{\alpha}^{2} \;
- \; \frac{\kappa^{2}}{24\beta}\sum_{\alpha=1}^{M} (n_{\alpha}^{3}-n_{\alpha})
\end{equation}
As the wave functions $\Psi_{\bf q, n}^{(M)}({\bf x})$ can be proved to constitute the complete 
and othonormal set, the replica partition function of the original 
directed polymer problem can be represented in the form 
of their linear combination:
\begin{equation}
\label{bas4-1}
\Psi({\bf x},t) \; = \; \sum_{M=1}^{N} \; 
\Biggl(\prod_{\alpha=1}^{M} \int_{-\infty}^{+\infty} \frac{dq_{\alpha}}{2\pi} \; \; 
\sum_{n_{\alpha=1}}^{\infty} \Biggr) \; 
 \Psi_{\bf q, n}^{(M)}({\bf x}) \Psi_{\bf q, n}^{(M)^{*}}({\bf 0}) \; 
\mbox{\LARGE e}^{-E_{M}({\bf q,n}) t }
\;{\boldsymbol \delta}\biggl(\sum_{\alpha=1}^{M} n_{\alpha}, \; N \biggr)
\end{equation}
where ${\boldsymbol \delta}(k, \; m)$ is the Kronecker symbol (which allows to extend the summation
over $n_{\alpha}$'s to infinity).
Using the explicit form of the wave functions $\Psi_{\bf q, n}^{(M)}({\bf x})$ 
the above expression (after somewhat painful algebra) reduces to (see \cite{BAS} for details)
\begin{equation}
   \label{bas4-12}
 Z(N,L) \; =  \mbox{\LARGE e}^{-\beta N L f_{0}} \;\;
\tilde{Z}(N,\lambda) 
\end{equation}
where $f_{0} = \frac{1}{24}\beta^4 u^2 - \frac{1}{\beta L} \ln(\beta^{3} u)$ 
is the linear (selfaveraging) free energy density, and
\begin{eqnarray}
\label{bas4-17}
\tilde{Z}(N,\lambda) &=& N! 
\int\int_{-\infty}^{+\infty} \frac{dy dp}{4\pi\lambda N} 
\; \Ai(y+p^{2}) \; \mbox{\LARGE e}^{\lambda N y}  + 
\\
\nonumber
 &+& N! \sum_{M=2}^{N} \frac{1}{M!} 
\biggl[\prod_{\alpha=1}^{M} \sum_{n_{\alpha}=1}^{\infty} \int\int_{-\infty}^{+\infty}
 \frac{dy_{\alpha} dp_{\alpha}}{4\pi\lambda n_{\alpha}} 
\Ai(y_{\alpha}+p^{2}_{\alpha}) \mbox{\LARGE e}^{\lambda n_{\alpha} y_{\alpha}}\biggr] 
\prod_{\alpha<\beta}^{M}
\frac{\big|\lambda(n_{\alpha}-n_{\beta}) -i(p_{\alpha}-p_{\beta})\big|^{2}}{
      \big|\lambda(n_{\alpha}+n_{\beta}) -i(p_{\alpha}-p_{\beta}) \big|^{2}} \;
{\boldsymbol \delta}\biggl(\sum_{\alpha=1}^{M}n_{\alpha}, \; N\biggr)
\end{eqnarray}
where $Ai(t)$ is the Airy function, and instead of the system length $L$ we have introduced a new parameter
$\lambda = \frac{1}{2} (\beta^{5} u^{2} L )^{1/3}$.
The first term in the above expression is the contribution of the ground state $(M=1)$,
and the next terms $(M \geq 2)$ are the contributions of the rest of the energy
spectrum. 

Using the Cauchy double alternant identity
\begin{equation}
 \label{tw4}
\frac{\prod_{\alpha<\beta}^{M} (a_{\alpha} - a_{\beta})(b_{\alpha} - b_{\beta})}{
     \prod_{\alpha,\beta=1}^{M} (a_{\alpha} - b_{\beta})} \; = \; 
(-1)^{N(N-1)/2} \det\Bigl[\frac{1}{a_{\alpha}-b_{\beta}}\Bigr]_{\alpha,\beta=1,...,M}
\end{equation}
the product term in eq.(\ref{bas4-17}) can be represented in the determinant form:
\begin{equation}
 \label{tw5}
\prod_{\alpha<\beta}^{M}
\frac{\big|\lambda(n_{\alpha}-n_{\beta}) -i(p_{\alpha}-p_{\beta})\big|^{2}}{
      \big|\lambda(n_{\alpha}+n_{\beta}) -i(p_{\alpha}-p_{\beta}) \big|^{2}}
\; = \; 
\prod_{\alpha=1}^{M} (2\lambda n_{\alpha}) \; 
\det\Bigl[\frac{1}{\lambda n_{\alpha} -ip_{\alpha} + \lambda n_{\beta} + ip_{\beta}}\Bigr]_{\alpha,\beta=1,...,M}
\end{equation}
Substituting now the expression for the replica partition function $\tilde{Z}(N,\lambda)$
into the definition of the probability function, eq.(\ref{tw3}), we can perform summation over $N$
(which would lift the constraint $\sum_{\alpha=1}^{M}n_{\alpha} =  N$) and obtain:
\begin{equation}
 \label{tw6}
W(x)  = 
\lim_{\lambda\to\infty} \Biggl\{
1  +  \sum_{M=1}^{\infty} \frac{(-1)^{M}}{M!}
\Biggl[\prod_{\alpha=1}^{M}
\int\int_{-\infty}^{+\infty}
\frac{dy_{\alpha} dp_{\alpha}}{2\pi}  
\Ai(y_{\alpha}+p^{2}_{\alpha})
\sum_{n_{\alpha}=1}^{\infty} (-1)^{n_{\alpha}-1} \mbox{\LARGE e}^{\lambda n_{\alpha} (y_{\alpha}+x)}
\Biggr] 
\det\Bigl[\frac{1}{\lambda n_{\alpha} -ip_{\alpha} + \lambda n_{\beta} + ip_{\beta}}\Bigr] \Biggr\}
\end{equation}
The above expression in nothing else but the expansion of the Fredholm determinant $\det(1 - \hat{K})$
(see e.g. \cite{Mehta})
with the kernel
\begin{eqnarray}
\nonumber
\hat{K} &\equiv&
K\bigl[(n,p); (n',p')\bigr] 
\\
&=&
\Biggl[\int_{-\infty}^{+\infty} dy \Ai(y+p^{2}) (-1)^{n-1} \mbox{\LARGE e}^{\lambda n (y+x)}\Biggr]
\frac{1}{\lambda n -ip + \lambda n' + ip'} 
\Biggl[\int_{-\infty}^{+\infty} dy \Ai(y+{p'}^{2}) (-1)^{n'-1} \mbox{\LARGE e}^{\lambda n' (y+x)}\Biggr]
\label{tw7}
\end{eqnarray}
Using the exponential representation of this determinant we get
\begin{equation}
 \label{tw8}
W(x)  = 
\lim_{\lambda\to\infty}
\exp\Bigl[-\sum_{M=1}^{\infty} \frac{1}{M} \; Tr \hat{K}^{M} \Bigr]
\end{equation}
where 
\begin{eqnarray}
\nonumber
Tr \hat{K}^{M} &=& 
\Biggl[\prod_{\alpha=1}^{M}
\int\int_{-\infty}^{+\infty}
\frac{dy_{\alpha} dp_{\alpha}}{2\pi}  
\Ai(y_{\alpha}+p^{2}_{\alpha})
\sum_{n_{\alpha}=1}^{\infty} (-1)^{n_{\alpha}-1} \mbox{\LARGE e}^{\lambda n_{\alpha} (y_{\alpha}+x)}
\Biggr] \times
\\
\nonumber
\\
&\times&
\frac{1}{(\lambda n_{1} -ip_{1} + \lambda n_{2} + ip_{2}) 
(\lambda n_{2} -ip_{2} + \lambda n_{3} + ip_{3}) ...
(\lambda n_{M} -ip_{M} + \lambda n_{1} + ip_{1})}
 \label{tw9}
\end{eqnarray}
Substituting here
\begin{equation}
 \label{tw10}
\frac{1}{\lambda n_{\alpha} -ip_{\alpha} + \lambda n_{\alpha+1} + ip_{\alpha+1}} \; = \; 
\int_{0}^{\infty} d\omega_{\alpha} 
\exp\bigl[-( \lambda n_{\alpha} - ip_{\alpha} + \lambda n_{\alpha+1} + ip_{\alpha+1}) \omega_{\alpha} \bigr]
\end{equation}
one can easily perform the summation over $n_{\alpha}$'s. Taking into account that
\begin{equation}
 \label{tw11}
\lim_{\lambda\to\infty} 
\sum_{n=1}^{\infty} (-1)^{n-1} \mbox{\LARGE e}^{\lambda n z} \; = \; 
\lim_{\lambda\to\infty} \frac{\mbox{\LARGE e}^{\lambda z}}{1 + \mbox{\LARGE e}^{\lambda z}} \; = \; 
\theta(z)
\end{equation}
and shifting the integration parameters, $y_{\alpha} \to y_{\alpha} - x + \omega_{\alpha} + \omega_{\alpha-1}$
and $\omega_{\alpha} \to  \omega_{\alpha} + x/2$, we obtain
\begin{equation}
 \label{tw12}
\lim_{\lambda\to\infty} Tr \hat{K}^{M} \; = \; 
\prod_{\alpha=1}^{M}
\int_{0}^{\infty} dy_{\alpha} 
\int_{-\infty}^{+\infty} \frac{dp_{\alpha}}{2\pi}  
\int_{-x/2}^{\infty} d\omega_{\alpha} \;
\Ai(y_{\alpha}+p^{2}_{\alpha}+\omega_{\alpha}+\omega_{\alpha-1}) \; 
\mbox{\LARGE e}^{ip_{\alpha} (\omega_{\alpha}-\omega_{\alpha-1})}
\end{equation}
where by definition it is assumed that $\omega_{0} \equiv \omega_{M}$. Using the Airy function integral representation,
and taking into account that it satisfies the differential equation, $\Ai''(t) = t \Ai(t)$, 
one can easily perform the following integrations:
\begin{eqnarray}
\label{tw13}
\int_{0}^{\infty} dy 
\int_{-\infty}^{+\infty} \frac{dp}{2\pi}  
\Ai(y + p^{2} + \omega + \omega') 
\mbox{\LARGE e}^{ip (\omega-\omega')} &=&
 2^{-1/3} \int_{0}^{\infty} dy 
\Ai\bigl(2^{1/3} \omega + y\bigr) 
\Ai\bigl(2^{1/3} \omega' + y\bigr) 
\\
\nonumber
&=&
\frac{\Ai\bigl(2^{1/3} \omega\bigr) \Ai'\bigl(2^{1/3} \omega'\bigr) - 
      \Ai'\bigl(2^{1/3} \omega\bigr) \Ai\bigl(2^{1/3} \omega'\bigr)}{
\omega - \omega'}
\end{eqnarray}
Redefining $\omega_{\alpha} \to \omega_{\alpha} 2^{-1/3}$ we find
\begin{equation}
\label{tw14}
\lim_{\lambda\to\infty} Tr \hat{K}^{M} \; = \; 
\int\int ...\int_{-x/2^{2/3}}^{\infty} d\omega_{1} d\omega_{2} ... d\omega_{M} \; 
K^{*}(\omega_{1},\omega_{2}) K^{*}(\omega_{2},\omega_{3}) ... K^{*}(\omega_{M},\omega_{1})
\end{equation}
where
\begin{equation}
 \label{tw15}
K^{*}(\omega,\omega') \; = \; 
\frac{\Ai(\omega) \Ai'(\omega') - \Ai'(\omega) \Ai(\omega')}{
\omega - \omega'}
\end{equation}
is the so called Airy kernel. This proves that in the thermodynamic limit, $L \to \infty$,
the probability function $W(x)$, eq.(\ref{tw1}), is defined by the Fredholm determinant,
\begin{equation}
 \label{tw16}
W(x)  \; = \; \det[1 - \hat{K}^{*}] \; \equiv \; F_{2}(-x/2^{2/3})
\end{equation}
where $\hat{K}^{*}$ is the integral operator on $[-x/2^{2/3}, \infty)$ with the 
Airy kernel, eq.(\ref{tw15}). The function $F_{2}(s)$ is the Tracy-Widom distribution \cite{Tracy-Widom}
\begin{equation}
 \label{tw17}
F_{2}(s) \; = \; \exp\Bigl(-\int_{s}^{\infty} dt \; (s-t) \; q^{2}(t)\Bigr)
\end{equation}
where the function $q(t)$ is the solution of the Panlev\'e II equation, $q'' = t q + 2 q^{3}$
with the boundary condition, $q(t\to +\infty) \sim Ai(t)$. 
This distribution was originally derived as the probability distribution of the
largest eigenvalue of a $n\times n$ random hermitian matrix in the limit $n \to \infty$.
At present there are exists an appreciable  list of statistical systems 
(which are not always look similar)
in which the fluctuations of the quantities which play the role
of "energy"  are described by {\it the same} distribution function
$F_{2}(s)$. These systems are: 
the polynuclear growth (PNG) model \cite{PNG_Spohn},
the longest increasing subsequences (LIS) model \cite{LIS}, 
the longest common subsequences (LCS) \cite{LCS},
the oriented digital boiling model \cite{oriented_boiling}, 
the ballistic decomposition model \cite{ballistic_decomposition}, 
and finally the zero-temperature  lattice version of the directed polymers 
with a specific (non-Gaussian) site-disorder distribution \cite{DP_johansson}.
Now we can add to this list one-dimensional directed polymers with Gaussian
$\delta$-correlated random potential.

%%%%%%%%%%%%%%%%%%%%%%%%%%%%%%%%%%%%%%%%%%%%%%%%%%%%%%%%%%%%%%%%%%%%%%%%%%%%%%%%%%%%%%%%%%%%%%%%%%%%
%%%%%%%%%%%%%%%%%%%%%%%%%%%%%%%%%%%%%%%%%%%%%%%%%%%%%%%%%%%%%%%%%%%%%%%%%%%%%%%%%%%%%%%%%%%%%%%%%%%%

%\vspace{20mm}

%\newpage

\end{document}